\documentclass[twocolumn,floatfix,groupedaddress,superscriptaddress,amsmath,a4paper,twoside,showkeys]{revtex4}
\pdfoutput=1
\usepackage{graphicx}
\usepackage{chemarr}
\usepackage{times,longtable}
\usepackage{hyperref,amsmath}
\usepackage{version}
\usepackage{color,pdflscape}
\usepackage{soul,amssymb}

\usepackage{epsf,mathtools}
\usepackage[usenames,dvipsnames]{xcolor}

\graphicspath{{figures/},{figures/pdf/},{figures/eps/}}

\bibpunct{[}{]}{,}{n}{}{,}

\providecommand{\addhyphen}[1]{#1.---}

\makeatletter
\renewcommand \paragraph{%
  \@startsection
    {paragraph}%
    {4}%
    {\parindent}%
    {\z@}%
    {-.1em}%
    {\normalfont\normalsize\itshape\addhyphen}%
}%
\makeatother

\begin{document}

\title{Counting and correcting thermodynamically infeasible flux cycles in genome-scale metabolic networks}
 

\author{D. De Martino}

\affiliation{Dipartimento di Fisica, Sapienza Universit\`a di Roma, p.le A. Moro 2, 00185 Roma (Italy)}

\affiliation{Center for Life Nano Science@Sapienza, Istituto Italiano di Tecnologia, Viale Regina Elena 291, 00161 Roma (Italy)}

\author{F. Capuani}

\affiliation{Dipartimento di Fisica, Sapienza Universit\`a di Roma, p.le A. Moro 2, 00185 Roma (Italy)}

\author{M. Mori}

\affiliation{Dipartimento di Fisica, Sapienza Universit\`a di Roma, p.le A. Moro 2, 00185 Roma (Italy)}

\author{A. De Martino}

\affiliation{Dipartimento di Fisica, Sapienza Universit\`a di Roma, p.le A. Moro 2, 00185 Roma (Italy)}

\affiliation{Center for Life Nano Science@Sapienza, Istituto Italiano di Tecnologia, Viale Regina Elena 291, 00161 Roma (Italy)}

\affiliation{IPCF-CNR, Unit\`a di Roma-Sapienza, Roma (Italy)}

\affiliation{{\rm Authors contributed equally}}

\author{E. Marinari}

\affiliation{Dipartimento di Fisica, Sapienza Universit\`a di Roma, p.le A. Moro 2, 00185 Roma (Italy)}

\affiliation{Center for Life Nano Science@Sapienza, Istituto Italiano di Tecnologia, Viale Regina Elena 291, 00161 Roma (Italy)}

\affiliation{{\rm Authors contributed equally}}

\begin{abstract}
Thermodynamics constrains the flow of matter in a reaction network to occur through routes along which the Gibbs energy decreases, implying that viable steady-state flux patterns should be void of closed reaction cycles. Identifying and removing cycles in large reaction networks can unfortunately be a highly challenging task from a computational viewpoint. We propose here a method that accomplishes it by combining a relaxation algorithm and a Monte Carlo procedure to detect loops, with {\it ad hoc} rules (discussed in detail) to eliminate them. As test cases, we tackle (a) the problem of identifying infeasible cycles in the {\it E. coli} metabolic network and (b) the problem of correcting thermodynamic infeasibilities in the Flux-Balance-Analysis solutions for 15 human cell-type specific metabolic networks. Results for (a) are compared with previous analyses of the same issue, while results for (b) are weighed against alternative methods to retrieve thermodynamically viable flux patterns based on minimizing specific global quantities. Our method on one hand outperforms previous techniques and, on the other, corrects loopy solutions to Flux Balance Analysis. As a byproduct, it also turns out to be able to reveal possible inconsistencies in model reconstructions. 
\end{abstract}

\keywords{thermodynamics; infeasible cycles; genome-scale metabolic networks; flux-balance analysis}

\maketitle

\tableofcontents

\section{Introduction}

Starting from the discovery by Lavoisier concerning the relation between respiration and combustion, 
thermodynamics stands as a key physical framework for  understanding metabolism and physiology, from single cell to whole organisms. When applied to a given metabolic reaction network, at the simplest level, thermodynamics  requires that, in non-equilibrium steady states, fluxes of matter proceed downhill in the underlying Gibbs (free) energy landscape. Violations of this rule (which corresponds to nothing but the second law of thermodynamics) are signaled by the existence of unphysical cycles in flux configurations \cite{Price:2002mw}. In the current era of metabolic genome-scale reconstructed networks, the implementation of such a constraint in computational models of a cell's metabolism has far-reaching implications \cite{nethatzi}, ranging from the physical feasibility of flux configurations \cite{babson} to the estimation of metabolite levels \cite{Hoppe:2007cr}, the assignment of directionality for reactions and pathways \cite{Qian:2005ve},  and the characterization of the overall chemical energy balance \cite{Beard:2002vn}. Accounting for thermodynamics in genome-scale models, however, poses considerable practical problems both for algorithms and for CPU costs. 

The reference modeling scheme that we shall consider here is given by the so-called constraint-based models \cite{palssonbook}, widely employed in the literature to descibe the operation of a biochemical reaction network at steady states with time-independent metabolite levels. While building a detailed model of metabolism presupposes knowledge of the kinetic parameters and reaction mechanisms \cite{AthelCBbook}, and should possibly take into
account stochasticity \cite{stocqian} and spatial diffusion \cite{brownianbio, Beg:2007mq}, constraint-based models focus on well-mixed non-equilibrium steady states (NESSs) for the reaction fluxes, to recover which the fundamental information comes from the underlying stochiometry alone. For a given stoichiometric matrix $\mathbf{S}=\{S_{mr}\}$ that accounts for the stoichiometric coefficient of metabolite $m$ in reaction $r$ (with the usual sign convention to distinguish products from substrates), a flux vector $\mathbf{v}=\{v_r\}$ represents a non-equilibrium steady state if it enforces the balance of metabolite levels $\mathbf{c}=\{c_m\}$, i.e. if
\begin{equation}\label{eq:fba}
\dot{\mathbf{c}} \equiv \mathbf{Sv}=\mathbf{0}~~.
\end{equation}
In usual applications, physiological aspects constrain fluxes to vary with certain ranges, so that  bounds of the type $v_r \in [v_{r}^{{\rm min}},v_{r}^{{\rm max}}]$ are normally prescribed for every reaction $r$. Such bounds may reflect, for instance, the fact that certain processes are known to be physiologically irreversible (e.g. $v_r \ge 0 $) or are required to occur at precise rates (as can be the case for maintenance reactions). From a geometric point of view, under (\ref{eq:fba}) and the bounds on fluxes, the space of possible NESSs is represented by a convex polytope. If all flux configurations inside this volume could be considered as physically realizable solutions, one might assess the `typical' productive capabilites of the network by sampling them using a controlled algorithm \cite{Kyoto10}. Unluckily this route often turns out to be computationally too expensive for large enough systems. Alternatively, one may search for the state(s) that maximize the value of certain biologically motivated objective functions, which can usually be cast in the form of a linear combination of fluxes that represents the selective production of a given set of metabolites. The flux configurations that maximize such a linear functional can be retrieved with the methods of linear programming \cite{linearpro}, the textbook case being growth yield maximization for bacterial cells in culture. Such a framework, known as Flux Balance Analysis (FBA) \cite{Orth:2010if}, has been shown to be predictive in many instances, even under genetic and/or environmental perturbations \cite{Segre:2002oq} (possibly with small modifications). 

Solutions of (\ref{eq:fba}) are in general not guaranteed to be thermodynamically viable. Frameworks like FBA can however be modified to include thermodynamic constraints directly in order to generate thermodynamically viable flux configurations, for instance by resorting to empirical data to estimate the chemical potentials of metabolites \cite{Jankowski:2008bh} and infer reaction reversibility more precisely \cite{Fleming:2009qa,Kummel:2006bh}. As a matter of fact, a large part of thermodynamic inconsistencies appear to be due to fallacious direction assignments. Models of this type, however, require prior biochemical information that is often scarce or unavailable \cite{Albertybook}.
To overcome these difficulties, new methods were devised that detect infeasible loops leveraging only on the constraint based model, i.e. on the structure on the metabolic network alone \cite{Schellenberger:2011uq,Henry:2007xz,fastthermo}. Although these methods release the need for experimental knowledge, the direct detection of infeasible loops is a computationally demanding task that limits their applicability.
It is therefore important to devise algorithms that are able to identify and remove thermodynamic inconsistencies from solutions of \eqref{eq:fba} or, more generally, from generic flux patterns.

Checking thermodynamic feasibility of a flux pattern can be made straightforward. Denoting by $\mathbf{v}'$ a flux vector, from which we exclude uptakes and every reaction that cannot be associated directly to a thermodynamic constraint (like biomass production, ``effective'' reactions with non-integer stoichiometry, or null fluxes), let us define the matrix $\boldsymbol{\Omega}=\{\Omega_{mr}\}$ with elements $\Omega_{mr}=-{\rm sign}(v'_r)S_{mr}$. Note that the minus sign in the definition of $\boldsymbol{\Omega}$ is needed to connect the directions of the reactions to the corresponding Gibbs energy differences. Thermodynamic feasibility of $\mathbf{v}'$ is easily seen to be guaranteed (see the Supporting Text for a toy example) if a non-zero vector $\boldsymbol{\mu}=\{\mu_m\}$ (of chemical potentials) exists such that \cite{chembiop}
\begin{equation}\label{eq:chempot}
\boldsymbol{\mu\Omega} > \mathbf{0}~~.
\end{equation}
(Note that, for physiologicaly realism, one would like the individual $\mu_m$'s to lie in specific ranges. As we shall not be concerned here with reconstructing the cellular Gibbs energy landscape \cite{noiplos}, this aspect will be neglected in what follows.) Solving \eqref{eq:chempot} can be done very efficiently, e.g., via relaxation algorithms \cite{linearpro}. By Gordan's theorem of the alternatives (see e.g. \cite{deMartino12}), if \eqref{eq:chempot} has no solution, then necessarily its dual system
\begin{equation}\label{eq:loop}
\boldsymbol{\Omega} \mathbf{k}=\mathbf{0}
\end{equation}  
with $\mathbf{k}=\{k_r\}$  possesses at least one non-zero solution with $k_r\geq 0$ for each $r$. It is easy to understand that such vectors $\mathbf{k}$ represent closed cycles of reactions that could in principle be able to perform work without using free energy, contradicting the laws of thermodynamics (note however that the topology of such cycles may turn out to be remarkably complex, see e.g. \cite{noiplos}). The problem posed by thermodynamics can then be seen as that of identifying and removing such loops. 

Finding all cycles in a directed (bipartite) network is, at heart, an integer programming problem in the NP-hard class \cite{jnsn}, which suggests that using deterministic algorithms to find loops in large enough networks may be unwise. However, for networks in which reliable prior thermodynamic information is available the complexity of loop counting can be significantly reduced, and indeed in some cases the problem has already been tackled (altough, in our view, not fully solved) in genome-scale networks with some degree of success \cite{noiplos,wagner}.  By contrast, in large networks lacking  detailed thermodynamic information, like the human metabolic networks, the implementation of thermodynamic constraints requires the development of algorithms that are able to handle more difficult instances of the loop counting problem. Luckily, in many hard computational problems where the use of exact algorithms is prevented by CPU costs, stochastic methods have proved to be effective. A biologically relevant case is represented by the problem of sampling solutions of (\ref{eq:fba}), for which Monte Carlo \cite{Wiback:2004kc,Price:2004lp} and message-passing techniques \cite{mezmon} are being employed instead of deterministic methods (as the latter presuppose the enumeration of a possibly exponential number of vertices of the polytope). It is simple to guess that a similar strategy might be employed (as we shall see, with some care) for the analysis of the solutions of (\ref{eq:loop}), i.e. to identify reaction cycles.

The strategy we present here combines a relaxation algorithm and a Monte Carlo method to allow for the thorough  analysis of thermodynamic infeasibilities on genome-scale metabolic networks of unprecedented size. More precisely, loops will be found by applying Monte Carlo to \eqref{eq:loop} with a reduced search space obtained by analysing how relaxation behaves when applied to (\ref{eq:chempot}). Once a loop is found, it can be removed in several ways, provided they don't violate any  of the constraints other than thermodynamic (e.g. mass balance). We shall discuss and compare different approaches: more precisely, a `local' rule that exploits, in essence, the fact that fluxes in cycles are defined up to a constant, and a `global' rule, based on the minimization of an overall function of the fluxes. The method will be used to analyze different types of networks of large size. Specifically, we shall first identify all loops in the metabolic network of {\it E. coli} \cite{Feist:2007zk}, then focus on amending the FBA solutions of 15 different human metabolic network models derived from the genome-scale reactome Recon-2 \cite{Recon2}, all bearing a specified objective function. Such solutions turn out to be rich with infeasible cycles, which we are able to find and correct. 

The structure and rationale of the method we propose are discussed in detail in Section 2, together with a brief summary of the network reconstructions we shall employ. Section 3 exposes our results, while our conclusions are reported in Section 4.

\section{Materials and methods}


\subsection{Materials: metabolic network reconstructions}

The human reactome Recon-2 \cite{Recon2} has been reconstructed by a community that merged and integrated existing global human metabolic networks and transcriptional information on specific human cell types. Authors verified the quality of Recon-2 by determining how many tasks the network was able to perform. A task can be as simple as the transformation of a metabolite by a single enzyme or by a complex pathway --like fermentation or oxydative phosphorylation-- or as complex as the production of the building blocks, energy, cofactors, etc. required for cell duplication, i.e. biomass. For \emph{E. Coli}, and in general for unicellular organisms, biomass yield is a valuable objective function for the FBA framework \cite{cobra2}, since its maximization essentialy equals growth maximization at fixed nutrient intake. Although it is unlikely that, in normal circumstances, cells in  a multicellular organism maximize the biomass yield, we stick to it as the FBA objective function as, for our purposes, the objective function can be seen merely as a tool to obtain motivated flux patterns for thermodynamic analysis.

In addition to the global reconstruction,  \cite{Recon2} provides a collection of 65 drafts of cell-specific networks, which are derived from Recon-2 by means of an automatic procedure that utilizes proteomic data \cite{cobra2,shlomi2008}. We focused on 15 networks with the ability to produce biomass, which we list in the first column of Table~\ref{TableFBA} together with the number of reactions ($N$) and metabolites ($M$) included in each case.  We have used in particular the network reconstructions in SBML format from \cite{Recon2} and resorted to the COBRA Toolbox \cite{cobra2} for the FBA analysis and to produce the Stoichiometric
matrix and the list of metabolites and reactions to be used in our analysis.

We have also analyzed the reconstructed metabolic network of the bacterium {\it E. coli} derived in \cite{Feist:2007zk}, consisting of 2382 reactions (including 305 uptakes) among 1668 metabolites. In this model, 548 reactions are putatively reversible. 

In each case, the key information we employed is encoded in the stoichiometric matrix $\mathbf{S}$.

\subsection{Methods}

\subsubsection{Algorithm for thermodynamic analysis: structure}

Before describing the algorithm in detail, we briefly recall the idea behind the procedure. We do not directly assess whether the flux configuration is loop free, but we try to compute the chemical potentials that satisfy Eq.~\eqref{eq:chempot}, which is a computationally easier problem to solve. If such a a solution exists, we are guaranteed that the flux configuration does not contain infeasible loops. It can be demonstrated~\cite{noiplos} that the relaxation method described below always converges polynomially to a solution, lack of convergence signals the presence of infeasible loops. Even when the relaxation method does not converge, it still provides us with a list of reactions that are likely to contain infeasible loops. To this limited subset of reactions, we can directly apply algorithms to find loop-free solutions.

\begin{figure}
\begin{center}
\includegraphics*[width=.48\textwidth,angle=0]{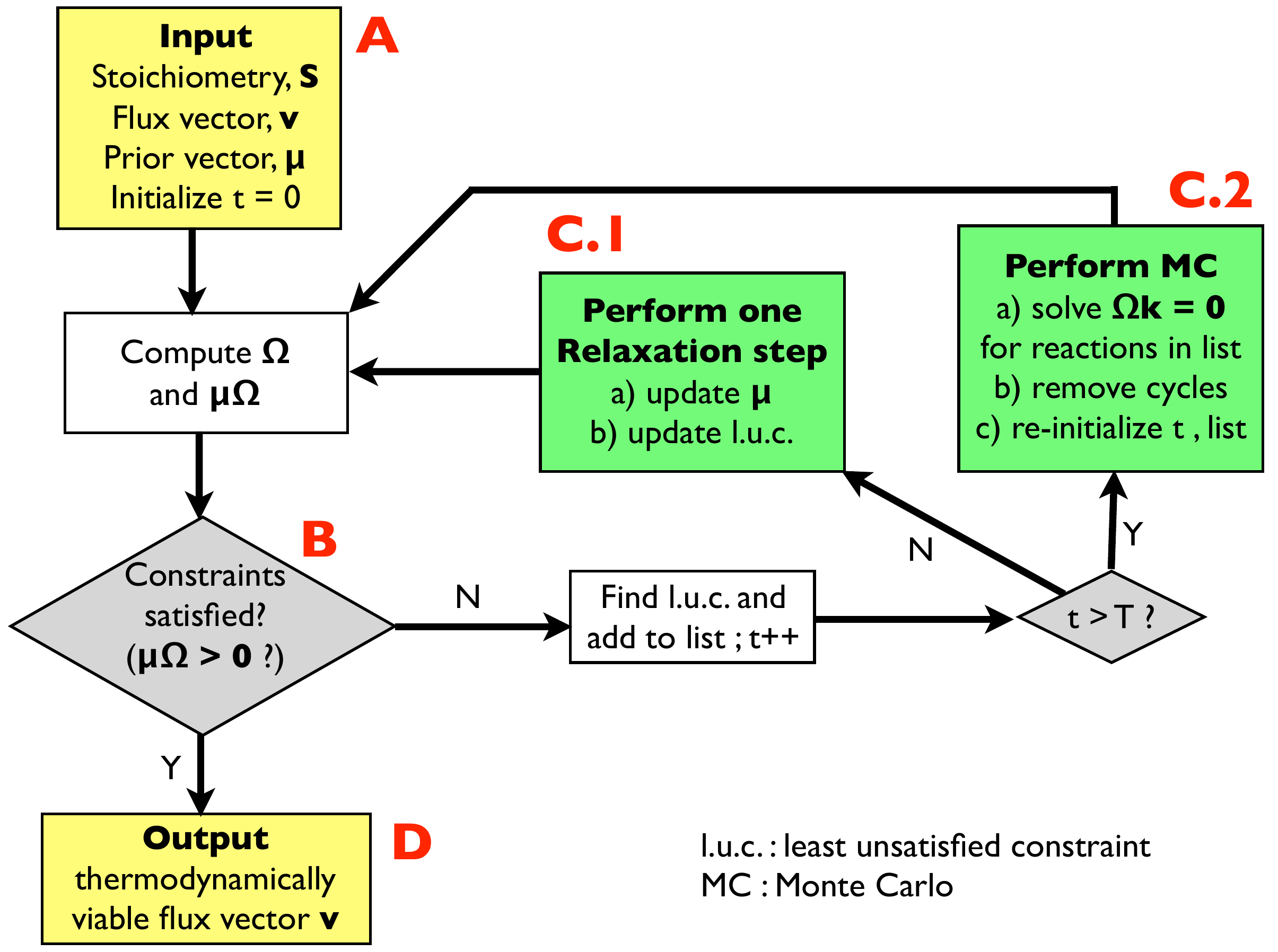}
\caption{\label{flow}Flowchart of the algorithm for counting and removing cycles employed in this study. See text for details.
}
\end{center}
\end{figure}

The overall structure of the algorithm is reported in Fig. \ref{flow}. 
In few words, and referring to the points A, B, C.1, C.2 and D shown explicitly in the flow chart:
\begin{itemize}
\item[(A)] Input: the input information includes a stoichiometric matrix $\mathbf{S}$, a flux vector $\mathbf{v}$ (e.g. a solution of FBA) and a prior vector $\boldsymbol{\mu}$ of chemical potentials. Initialize an integer variable $t$ at $t=0$ (relaxation steps) and an empty list.  
\item[(B)] Compute the matrix $\boldsymbol{\Omega}$  and evaluate the thermodynamic constraints (\ref{eq:chempot}), i.e. compute $\boldsymbol{\mu\Omega}$. If they are satisfied, i.e. if $\boldsymbol{\mu\Omega>0}$, go to (D); else register the least unsatisfied constraint (l.u.c.), i.e. the value of the index $r$ for which the corresponding components of the vector $\boldsymbol{\mu\Omega}$ is smallest (more negative), insert it into the list, and increase the $t$ variable by $1$ ; if $t\leq T$, with $T$ a pre-defined large parameter, go to (C.1); else go to (C.2).
\item[(C.1)] Update the vector $\boldsymbol{\mu}$ by performing a single step of the relaxation algorithm described in Sec. \ref{relax}; update the list by inserting the new l.u.c. and go back to (B). 
\item[(C.2)] Perform a Monte Carlo computation, as described in Sec. \ref{monte}, in order to find a solution of system \eqref{eq:loop}, namely $\boldsymbol{\Omega}\mathbf{k=0}$, including only the reactions appearing in the list. Once a solution is found, correct the associated cycle as described in Sec. \ref{correct1} and \ref{correct2}; re-initialize $t$, empty the list, and go back to (B).
\item[(D)] Output: a thermodynamically feasible flux vector.
\end{itemize}
 
In the following sections, we shall describe the sub-procedures (relaxation method, Monte Carlo and cycle removal) of the algorithm in detail. A C++ code which performs each of the above steps is provided as Supporting Material. It is worth to point out that the present study is not concerned with the calculation of realistic chemical potentials. Rather, we simply require their existence in order for the flux configuration to be feasible. In this case, the prior vector of chemical potentials can be arbitrary, e.g. constant or composed by i.i.d. random variables. If complemented with specific experimentally determined or computationally estimated priors for the chemical potentials, however, the relaxation method included in the above algorithm generates, as a by-product, a free energy vector compatible with the final flux configuration and can thus be employed to refine experimental data on the free energy of formation  of metabolites and/or on their levels \cite{noiplos}. Better priors ultimately allow to obtain more precise estimates for the real chemical potentials, but are essentially irrelevant for the convergence of the relaxation method. In what follows we shall neglect this aspect, which is discussed in depth in \cite{noiplos}, and focus exclusively on the retrieval of cycles.

\subsubsection{Checking thermodynamic viability by relaxation}\label{relax}

This routine, corresponding to point C.1 of the flow chart, allows to retrieve a solution of (\ref{eq:chempot}) starting from a vector of chemical potentials that is not a solution thereof. For simplicity, we construct an initial vector made of uniformly distributed random numbers. At any step $t$ of the procedure, given a chemical potential vector $\boldsymbol{\mu}(t)$, the relaxation algorithm corrects the l.u.c. of \eqref{eq:chempot} through the dynamics defined by
\begin{gather}
r_t ={\rm arg}~\min_r \sum_m \Omega_{mr} \mu _m(t) \\
\mu _m(t+1) = \mu_m(t) +\alpha \Omega_{mr_t}~~~~~\forall m~~,
\end{gather}  
where $r_t$ is the index of the l.u.c. added to the list at the $t$-th step and $\alpha>0$ is a constant. As explained in \cite{noiplos}, the above step simply shifts chemical potentials in a direction that will improve the l.u.c. The parameter $\alpha$ can be chosen in different ways, from a suitably small constant (as in the so-called Minover scheme \cite{minover0}) to a quantity proportional to the amount by which the constraint is violated (as in the Motzkin scheme \cite{linearpro}).
The Minover scheme returns a solution that is tipically closer to the prior, but here we utilized the faster Motzkin scheme. The above algorithm is known to converge to a solution of (\ref{eq:chempot}), upon iteration, in polynomial time if and only if a solution exists. If convergence fails, instead, by the Gordan theorem the reaction pattern contains infeasible cycles.  Convergence of relaxation therefore guarantees feasibility of a flux vector. In presence of loops, an iteration of the above dynamics tipically cycles among inconsistent constraints. Therefore, by keeping track of the l.u.c. over the iterations, i.e. by recording the the series $\{r_t\}$ for $t\geq 0$ (corresponding to the content of the list described in the flow chart), one can build a list of reactions that are candidates for being responsible for the failed convergence. If relaxation doesn't converge to a solution in a resonable time (denoted as $T$ above), we look for infeasible cycles, i.e. for solutions of \eqref{eq:loop},  within such a restricted list. To this aim, we employ a Monte Carlo method.

\subsubsection{Identifying loops by Monte Carlo} \label{monte}

As said above, cycles generically correspond to solutions of \eqref{eq:loop} with $\mathbf{k\geq 0}$. As the stoichiometric coefficients are typically integers, one can focus on searching solutions with $k_r$ non-negative integers for each $r$. To this aim, the following method (borrowed from the standard statistical physics toolbox) can be employed. Starting from \eqref{eq:loop}, note that the function
\begin{equation}\label{energy}
E (\mathbf{k}) =\sum_m \left(\sum_r \Omega_{mr}k_r\right)^2
\end{equation}
vanishes when $\mathbf{k}$ defines a flux cycle. Because $E\geq 0$, then, infeasible loops correspond to the minima of $E$, and loop finding amounts to locating the minima of $E$ in the search space $k_r\in \{0,1,2,\ldots\}$ for each $r$. Monte Carlo methods are ideally suited to tackle this type of problems \cite{montec}. In brief, such methods (the most famous of which is possibly the Metropolis scheme) generically generate vectors $\mathbf{k}$
distributed according to
\begin{equation}\label{boltz}
P(\mathbf{k}) \propto e^{-\beta E(\mathbf{k})}~~,
\end{equation} 
where $\beta>0$ is an externally fixed parameter. When $\beta\to\infty$, the above measure concentrates around the minima of $E$. One possibility to make sure that large enough values of $\beta$ are reached is to initialize the Monte Carlo simulation at some small value of $\beta$ and then increase $\beta$ in a controlled way, initializing each time from the configuration retrieved at the previous value of $\beta$ (`simulated annealing'). This is precisely the approach we have employed here:  in order to identify the infeasible loops, we have performed iterated Metropolis-based annealings to minimize the fictitious `energy'  \eqref{energy}. (The increase in performance warranted by  the annealing procedure compared to the simple Metropolis scheme at a fixed temperature is discussed in the Supporting Text.)

It is worth noting that, based on the above discussion of the relaxation method, the number of reactions to be included in the above procedure equals the number of distinct reactions appearing in the list, which is usually much smaller than $N$. To give an idea, in the study of {\it E. coli} whose results are reported below, our lists ended up containing at most 50 reactions, to be compared with the over 2000 that form the genome-scale reconstruction. Hence the computational costs of the Monte Carlo step of our algorithms are overall modest.

\subsubsection{Correcting the flux configuration: local strategy}\label{correct1}

Once a flux cycle has been identified, there are mutiple ways to remove it and re-organize the flux pattern while still preserving all constraints and, eventually, the values of objective functions.

\begin{figure}
 \centering
 \includegraphics*[width=0.48\textwidth]{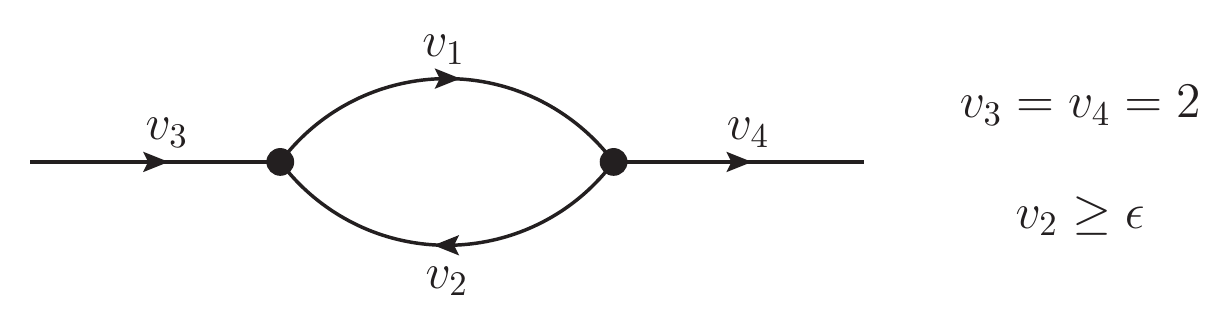}
 \caption{Example of a toy reaction network. The black dots are two metabolites, to each of which corresponds a mass balance constraint.
 Each line is labeled with the name of the flux carried by a reaction, and the arrow indicates the conventional forward direction of the fluxes.
 Evidently, a thermodynamically infeasible cycle is present if $v_1$ and $v_2$ have the same sign. \label{fig:example}}
\end{figure}

To clarify the situation, consider the following simple example with four reactions, pictured in Figure \ref{fig:example}.
$v_1$ and $v_2$ are ``internal'' fluxes, while $v_3$ and $v_4$ are an intake and an outtake flux, respectively.
The stoichiometric matrix $\mathbf{S}$ and the flux vector $\mathbf{v}$ reads:
\begin{align}
\mathbf{S}=\begin{pmatrix} -1 & 1 & 1&0\\ 1 & -1 &0& -1 \end{pmatrix} \label{eq:example_S_ext}
\qquad,\qquad \mathbf{v}=\begin{pmatrix} v_1 \\ v_2 \\ v_3 \\ v_4 \end{pmatrix} ~~.
\end{align}
With this formalism, the mass balance constraints \eqref{eq:fba} are given by the homogeneous equation $\mathbf{S}\mathbf{v}=\mathbf{0}$.
We also fix the value of the uptakes to, for example, $v_3=v_4=2$, and add a lower bound on the first flux, $v_2 \ge \epsilon$ (with $\epsilon$ a constant).

After the elimination of the uptakes, the internal stoichiometric matrix and flux vector (which for clarity in this section we denote as $\mathbf{S}^{int}$ and $\mathbf{v}^{int}$) are given by
\begin{align}
\mathbf{S}^{int}=\begin{pmatrix} -1 & 1 \\ 1 & -1  \end{pmatrix} \label{eq:example_S}
\qquad,\qquad\mathbf{v}^{int}=\begin{pmatrix} v_1 \\ v_2 \end{pmatrix} ~~.
\end{align}
Since we fixed the values of $v_3$ and $v_4$, we can move them to the r.h.s. of the mass balance constraints, obtaining
\begin{equation} \label{eq:FBA_reduced}
 \mathbf{S}^{int}\mathbf{v}^{int}=\mathbf{u} \qquad,\qquad \mathbf{u}=\begin{pmatrix} -v_3 \\ v_4 \end{pmatrix}~~.
\end{equation}
Equation \eqref{eq:FBA_reduced} must be solved together with the constraint $v_2\ge \epsilon$,
which we will treat separately.

Given a solution $\mathbf{v}$ of \eqref{eq:FBA_reduced}, any linear combination of the form
\begin{equation}\label{linearcomb}
\mathbf{v}'\equiv\mathbf{v}'(\mathbf{L})=\mathbf{v}+\sum_a L^a \mathbf{n}^a
\end{equation}
still satisfies the mass balance equations \eqref{eq:FBA_reduced}, provided
$\mathbf{n}^a$ is in the right null space of the internal stoichiometric matrix, i.e. is a solution of $\mathbf{S}^{int}\mathbf{n}^a=0$, and 
$\mathbf{L}=\{L^a\}$ is a family of real numbers.
If we now recall that loops $\mathbf{k}$ are non negative solutions of  Eq.~\eqref{eq:loop}, i.e. of $\mathbf{\Omega}\mathbf{k}=\mathbf{0}$,
where the matrix $\mathbf{\Omega}$  is built from $\mathbf{S}^{int}$ as $\Omega_{mr}=-\text{sign}(v_r)S^{int}_{mr}$,
we see that, by construction,  the relation $n_r^a=\text{sign}(v_r) k_r^a$  connects infeasible loops $\mathbf{k}^a$ ($a=1,2,\ldots$)
and the solutions of $\mathbf{S}^{int}\mathbf{n}^a=0$.
In other terms, the presence of cycles causes a degeneracy in flux patterns,
as there are many ways to assign fluxes to reactions in a cycle. 

Eq.~\eqref{linearcomb} can be used to correct the infeasible loops, while still satisfying all mass balance constraints.
The simplest possible correction scheme is based on the idea that by
properly fixing the value of the coefficients $L^a$ one can lift the degeneracy and rid the flux configuration of loops.
There is however a major caveat. To make it explicit, we note that (a) the signs of the fluxes $v'_r$ depend on the choice of
the coefficients $\mathbf{L}$, so that the matrix $\mathbf{\Omega}$ will also depend on it, and (b) it is not guaranteed that
the new fluxes $\mathbf{v}'$ will vary within the same bounds as $\mathbf{v}$.
This means that equation \eqref{eq:loop} must be solved together with all other constraints which are not related
to the stoichiometry, such as sign constraints. We will write these constraints in a general fashion as
$\mathbf{A} \mathbf{v}' \ge \mathbf{b}$. In our example, this matrix inequality reduces to $v_2 \ge \epsilon$.

With this notation, the space of vectors $\mathbf{L}$ yielding thermodynamically feasible solutions
is given by $\mathcal{C}_1\cap\mathcal{C}_2$, where
\begin{gather}
 \mathcal{C}_1 = \left\{ \mathbf{L} : \mathbf{A} \mathbf{v}' \ge \mathbf{b} \right\}\\
 \mathcal{C}_2 = \left\{ \mathbf{L} : \nexists \, \mathbf{k} \ge 0 ~~\text{such that}~~ \mathbf{\Omega(L)k}=0 \right\}
\end{gather}
The first set contains all constraints which are not related to stoichiometry, while
the second one contains the thermodynamic ones.
Now, unluckily, these two sets may or may not have points in common,
depending on the properties of the network and on the additional constraints
(in other words, it may not be possible to choose $\mathbf{L}$ properly).

Suppose now that, in our example, we are given the flux vector $\mathbf{v}^\star=(3,1)$ as solution to Eq.~\eqref{eq:FBA_reduced}.
We see that the vector $\mathbf{n}=(1,1)$ is in the null space of $\mathbf{S}^{int}$,
and that the new vector $\mathbf{v}'=\mathbf{v}^\star+L\mathbf{n}$ still satisfies (\ref{eq:FBA_reduced}).
Also, since both $v_1^\star$ and $v_2^\star$ are positive, $\mathbf{n}$ itself identifies a loop, i.e. it is a solution of \eqref{eq:loop} (for, in this case, $\mathbf{\Omega}=-\mathbf{S}^{int}$). It can be easily checked that, in this example, as long the constraint $v_2 \ge \epsilon$ is not taken into account we can pick $L$ in the interval $[-3,-1]$ to get rid of the cycle. The choice is arbitrary and produces a fully directional flux pattern such that reactions $v_1$ and $v_2$, if both active, operate in the same direction. The constraint $v_2 \ge \epsilon$, however, implies $L \ge \epsilon -1$. The sets $\mathcal{C}_1$ and $\mathcal{C}_2$ are then given by
\begin{gather}
 \mathcal{C}_1 = \left\{ L : L \ge \epsilon -1 \right\} \\
 \mathcal{C}_2 = \left\{ L : -3 \le L \le -1 \right\} 
\end{gather}
So we see that if $\epsilon<0$ there are infinitely many values of $L$ which remove the cycle, whereas the cycle cannot be removed if $\epsilon>0$. In other words, each time a flux is constrained to keep a pre-defined sign there is no guarantee that the loops involving this reaction can be corrected by simply lifting the flux degeneracy associated to them. The value $\epsilon=0$ plays here a particular role, since $v_2\ge0$ is an irreversibility constraint. In this case, the intersection of the two sets above imposes $L=-1$. This is indeed the most common kind of constraint on the fluxes, and allows for the simplest unambiguous loop removal strategy: setting the smallest flux to zero without changing signs to any of the other fluxes involved (as occurs in the present case upon choosing $L=-1$).

From this last observation we can deduce the following general loop removal strategy (to which we refer as the `local' correction strategy): for a given loop $\mathbf{k}^a$, choose the value of $L^a$ that sets to zero the flux of the reaction
whose absolute value is the smallest, i.e. 
\begin{equation}
L^a = - \min_{r~:~k^a_r >0} \frac{|v_r|}{k^a_r} ~~.
\end{equation}
With this choice, every constraint of the form $v_r\ge0$ will still be satisfied, and at least one loop will be removed. This strategy tends, in a sense, to minimize the distance between the original flux pattern and the corrected one. (We will quantify this aspect more precisely below.) On the other hand, if bounds like $v_r\ge \epsilon$ with $\epsilon>0$ are present, a more careful analysis is required.
We finish by noting that the most important instance of a constraint of the latter type in models of metabolism is represented by the ATP maintenance flux.

\subsubsection{Correcting the flux configuration: global strategy}\label{correct2}

Other possible loop-removing procedures are based on the minimization of some norm of the fluxes, as suggested in \cite{plos1}. Let us elaborate this idea further, and consider the function $Q_p(\mathbf{v})=\sum_r |v_r|^p$ with $p \ge 1$, representing, for different $p$'s, different norms of the flux vector $\mathbf{v}$ ($Q_1$ is the so-called `Taxicab' norm,  $Q_2$ is the square of the Euclidean norm, etc.). Suppose we have an FBA solution $\mathbf{v}^\star$ which minimizes $Q_p$. If $\{\mathbf{n}^a\}$ is the set of all null space vectors of the stoichiometric matrix, we can construct a new solution $\mathbf{v}'=\mathbf{v}^\star + \sum_a L^a \mathbf{n}^a$ and compute the partial derivative of $Q_p$ with respect to a coefficient $L^a$. The derivatives vanish when evaluated at $\mathbf{L=0}$:
\begin{multline}\label{deriva}
 \left. \frac{\partial}{\partial L^a} Q_p\left(\mathbf{v}^\star + \sum_b L^b \mathbf{n}^b\right) \right|_{\mathbf{L=0}} =\\
=  \sum_r p|v^\star_r|^{p-1} \text{sign}(v^\star_r) n^a_r =0~~.
\end{multline}
We see immediatly that the quantities $k_r^a = \text{sign}(v^\star_r) n^a_r$ can not have a definite sign. Restricting the sum to all terms with $n^a_r \neq 0$, we have two cases:
\begin{enumerate}
 \item If at least one of the fluxes $v^\star_r$ is zero, this reaction can not be involved in any cycle. In particular, $\mathbf{n}^a$ is not associated to a loop.
 \item If all fluxes are non-zero, the vector $\mathbf{k}^a$ cannot have a definite sign (positive or negative) since the sum of its entries, namely (\ref{deriva}), weighted with some positive coefficients, is zero. 
\end{enumerate} 
Therefore, the vector $\mathbf{v}^\star$ which minimizes $Q_p$ does not contain cycles.

The argument can be easily extended to include irreversibility constraints. Let $\mathbf{v}^\star$ denote the flux configuration which minimizes $Q_p(\mathbf{v})$ with irreversibility constraints $v_r \ge 0$ for the reactions $r$ beloging to the set $\mathcal{I}=\{r_1,~r_2,~\dots \}$. We shall instead denote by $\mathcal{I}_0=\{r_1^\prime,~r_2^\prime,~\dots\}\subseteq \mathcal{I}$ the set of irreversible reactions for which $v_r=0$ in $\mathbf{v}^\star$. Clearly, $\mathbf{v}^\star$ also minimizes $Q_p$ subject to the stronger constraints $v_r = 0$ for $r \in \mathcal{I}_0$ and $v_r>0$ for $r\in\mathcal{I}\setminus\mathcal{I}_0$. Given this, one can now proceed along the same lines as before, because, for any vector $\mathbf{n}^a$ in the null space of $\mathbf{S}^{int}$,
\begin{itemize}
 \item If some reaction $r$ for which $n^a_r\neq 0$ is forced to have zero flux since $r\in\mathcal{I}_0$, then $\mathbf{n}^a$ is not associated to a cycle;
 \item Otherwise, we can demonstrate that $\mathbf{n}^a$ does not correspond to a cycle by taking the partial derivative of $Q_p(\mathbf{v}^*+L^a \mathbf{n}^a)$ as done above.
\end{itemize}

Problems may arise, as before, when boundary conditions like $v_r\ge\epsilon$ with $\epsilon>0$ have to be considered. In particular, if the flux of a variable thus bounded is fixed to take the value $\epsilon$ there is the possibility that the cycle cannot be removed. In the example discussed above (see Figure \ref{fig:example}),
\begin{itemize}
 \item If $\epsilon < -1$, then $v_2 > \epsilon$ and the $Q_p$ minimization yields $\mathbf{v}^\star=(1,-1)$;
 \item If $-1 \le \epsilon \le0$, then $v_2=\epsilon$ but the flux configuration $\mathbf{v}^\star=(1-\epsilon,\epsilon)$ is still feasible (in particular, the configuration is feasible for $\epsilon=0$);
 \item If $\epsilon > 0$, then $v_2=\epsilon$ and the optimal flux configuration is not feasible.
\end{itemize}
In summary, the global minimization of the norm $Q_p$ produces thermodynamically feasible flux patterns, provided they are allowed by the constraints. If not, the minimization can get rid of all loops not involving reactions constrained to keep the same sign. We shall term the cycle-removal strategy based on minimizing a norm as the `global' strategy.

\section{Results}

\subsection{A test: identifying infeasible loops in the E. Coli network iAF1260}

As a proof of principle, we have applied our method to search and enumerate all independent infeasible loops of a large metabolic network reconstruction for the bacterium \emph{E. Coli}, the iAF1260 \cite{Feist:2007zk}. Other authors have attempted to solve the same enumeration problem before (see e.g. \cite{wagner}). We note however that our method is radically different in that we make use of the theorem of alternatives and do not directly search for loops on the graph, which is the more standard route \cite{Schilling:2000p4106}, or rely on subsequent optimizations that reduce the search space \cite{wagner, Mahadevan:2003pi}. In the present case, we characterize cycles in an ensemble of net-flux patterns generated randomly by assigning a specific operating direction to each reaction according to its reversibility. (More precisely, random flux patterns are generated by simply assigning an operation direction for each reaction as follows: if the reaction is irreversible, we pick the allowed direction; if the reaction is reversible, we select the forward or the reverse direction randomly with probability 1/2 (note that direction assignments suffice to pose the problem of thermodynamic feasibility). In this way, {\it all} reactions are active, a worst-case scenario with respect to a growth-yield optimizing state that normally only requires the operation of around 30\% of the reactions, implying (in our case) a much larger number of loops and, in principle, higher computational costs for loop counting. For each configuration, we look for and eliminate cycles until the material flow is thermodynamically consistent, recording the cycles that we have detected. Finally, we keep only independent loops by applying Gaussian elimination (i.e., we exclude from our list loops that can be decomposed as the sum of, say, two simpler loops).

\begin{figure}
\begin{center}
\includegraphics*[width=.48\textwidth,angle=0]{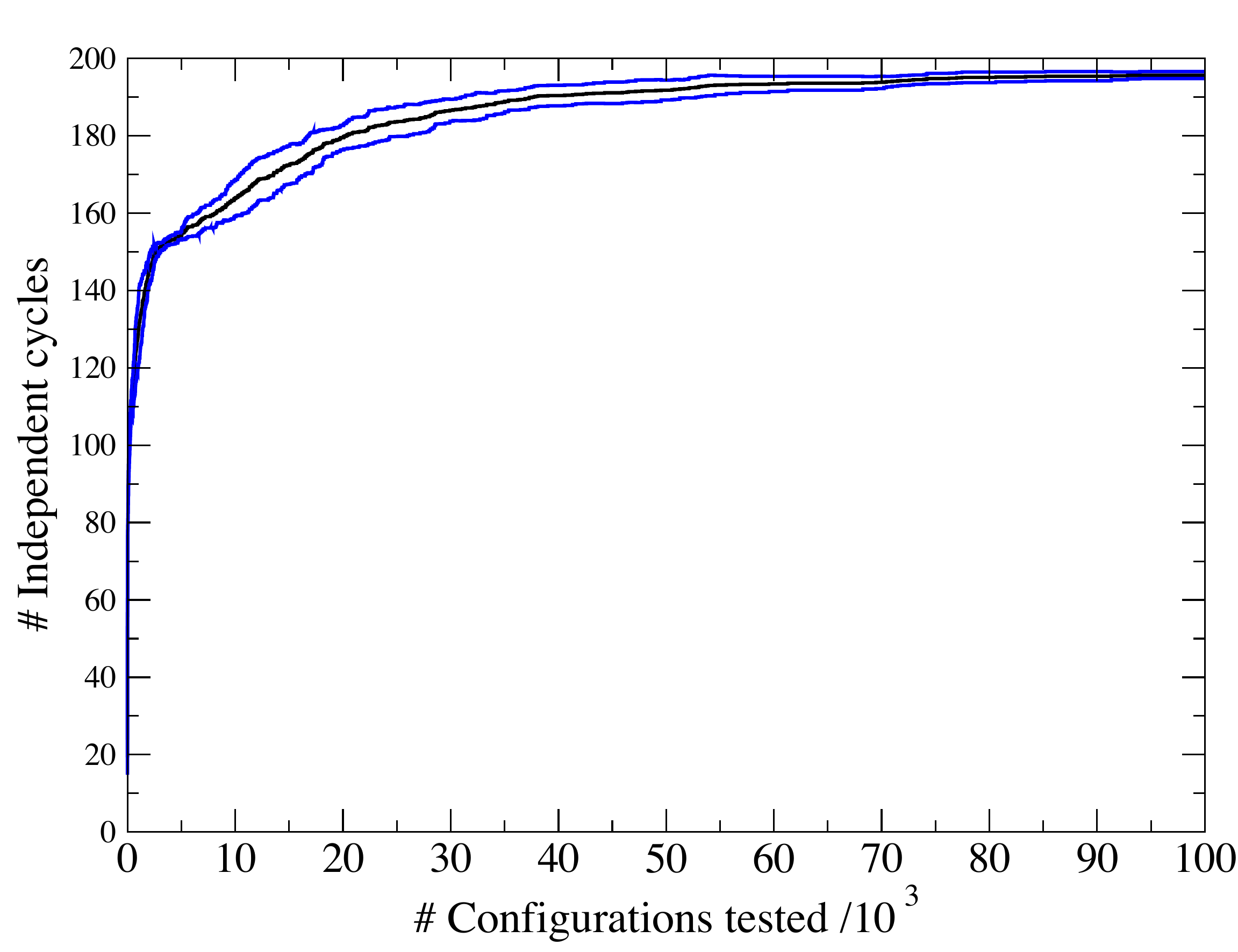}
\caption{Number of independent loops identified in the metabolic network of {\it E. Coli} iAF1260 as a function of the number of random configurations tested. Results are obtained by jackknifing over 10 configurations. The black line represents the average number of loops, while the blue lines represent the extremes of the error bars at each point.\label{fig2}
}
\end{center}
\end{figure}

\begin{figure}
\begin{center}
\includegraphics*[width=.48\textwidth,angle=0]{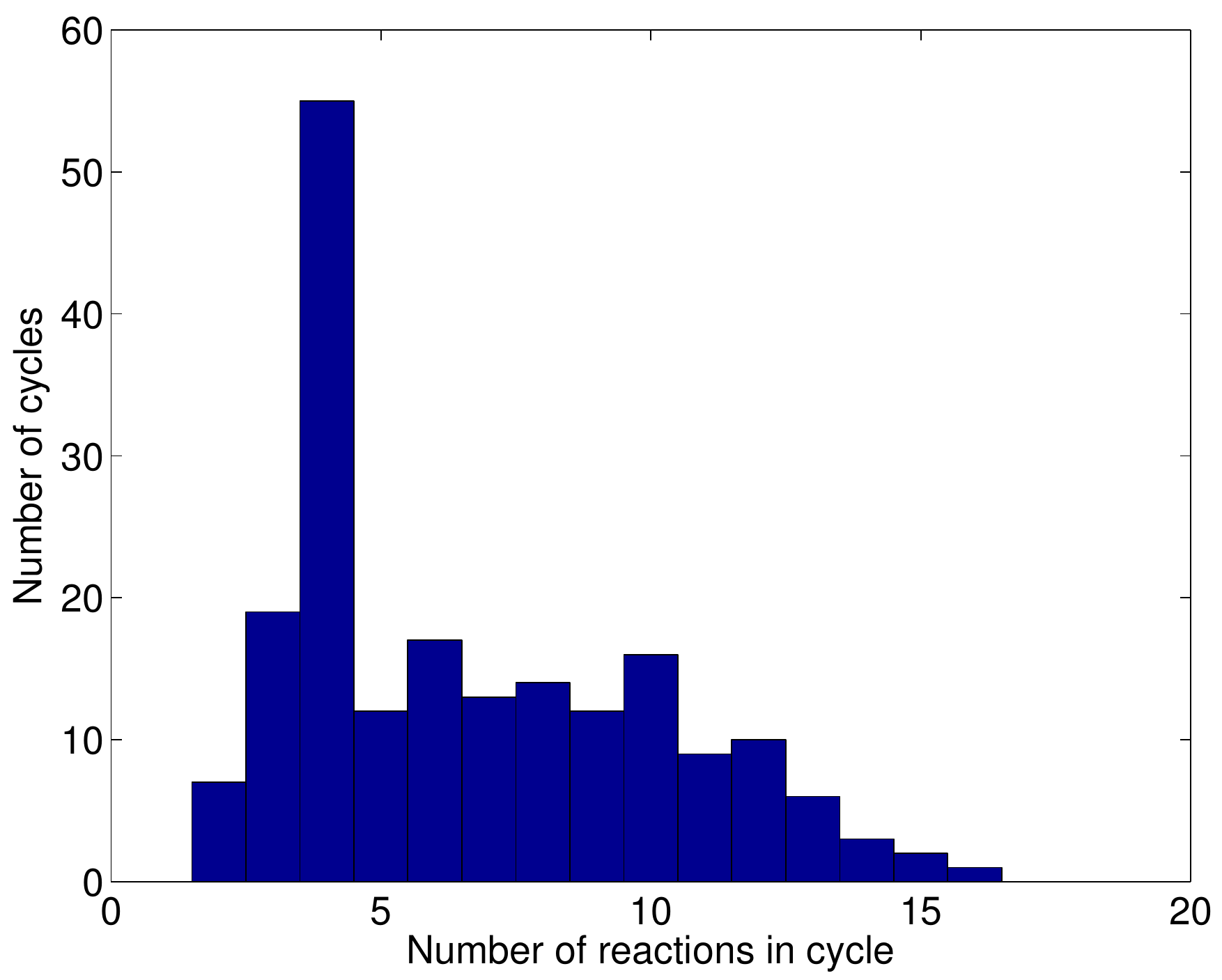}
\caption{Histogram of the length (number of reactions involved) of the 196 independent cycles detected in {\it E. Coli} iAF1260. \label{isto}}
\end{center}
\end{figure}

In Fig.~\ref{fig2}, we display the number of independent loops that we identify as a function of the number of random configurations tested. 
We identify 196 loops (189 of which turn out to be of size three or more) after having generated about 80,000 random configurations, and no new loops appear upon enlarging the test ensemble. The loops thus found are listed in the Supporting File 1, and a histogram of the cycle lengths (in terms of the number of reactions involved) is displayed in Figure~\ref{isto}.
We note that, in \cite{wagner}, 591 cycles were identified, 564 of which are however formed by 2 reactions, mostly originating from the fact that reversible reactions were, in that study, split in two separate processes (forward and reverse). Therefore, only 27 of those cycles were formed by 3 reactions or more. Because we do not split reversible reactions, we find only 7 cycles of length 2 and 189 loops of length at least equal to 3. We note that these 189 cycles span 396 reactions altogether. This suggests that, for the technique employed in \cite{wagner}, some loops were undetectable once the exhaustive search had been restricted to 50 reactions. We stress however that the procedure discussed in \cite{wagner} is in principle exact and once the restriction is removed it might be able to identify more cycles involving at least three reactions.

\subsection{Inconsistencies in the FBA solution for the overall human reactome Recon-2}

We now move on to the identification of thermodynamic infeasibilities in the human reactome Recon-2 \cite{Recon2}. In specific, we have analyzed the feasibility of flux patterns defined by solving FBA on the entire reactome, using the `biomass' reaction that comes with the reconstruction as the objective function. This section provides a concrete example of an inconsistency that is unrecoverable without correcting basic structural information concerning the network. It should be kept in mind, however, that the physiologically relevant metabolic networks that can be obtained from Recon-2 are the cell-type specific ones, which will be discussed in the following section.

\begin{figure}
\begin{center}
\includegraphics*[width=.48\textwidth,angle=0]{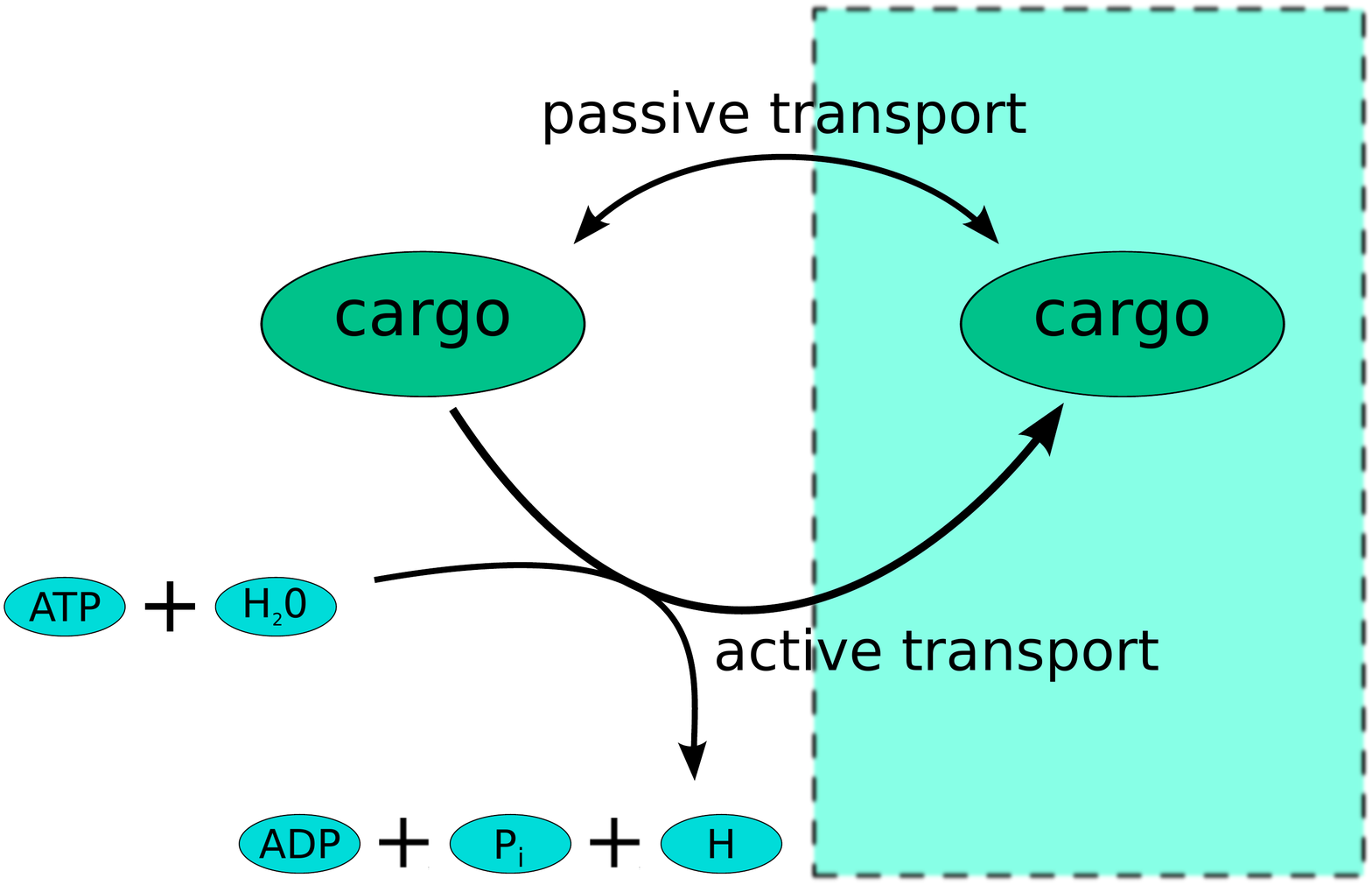}
\caption{Typical structure of an infeasible loop created by two reversible transports across a membrane-enclosed compartment: a passive one (by diffusion) and an active one (requiring the expenditure of energy). If the active process is allowed to reverse and the cargo re-enters the cell or compartment via diffusion, an infeasible loop is generated that builds ATP from ADP without energetic costs.\label{Fig:loopATP}}
\end{center}
\end{figure}

As almost all metabolic objective functions, the biomass reaction of Recon-2 contains ATP hydrolysis, representing the energetic requirements associated to cell duplication that are not explicitly accounted for by the flux organization.
As such requirements are typically large, the stoichiometry of ATP in the biomass reaction is often two orders of magnitude larger than that of the other chemical species. Hence, ATP tends to be the limiting factor for biomass production and FBA solutions will often organize metabolic fluxes so as to produce as much ATP as possible. This however turns out to lead, in Recon-2, to a violation of thermodynamics. In particular, in the FBA solution for Recon-2 we detect a huge number of cycles involving the active and passive transport of a metabolite through a membrane, as e.g. for the transport of Stearoyl-CoA (stcoa) from cytosol (c) to peroxisomes (x), namely (see Figure \ref{Fig:loopATP})
\begin{eqnarray} \label{ATPloop1}
\mbox{stcoa[c]} & \rightleftarrows &  \mbox{stcoa[x]} \\
\label{ATPloop2}
\mbox{adp[c] + h[c] + pi[c] + stcoa[x]} & \rightleftarrows  & \mbox{h2o[c] + atp[c] + stcoa[c]}\nonumber
\end{eqnarray}
Note that both reactions are listed as reversible. 
When the chemical potential difference drives stcoa from peroxisome to cytoplasm, the cell can actively transport Stearoyl-CoA to peroxisomes by consuming ATP. 
By reverting both reactions, however, the cell could produce ATP at no expense. This is precisely the type of solution that we obtain when we maximize the biomass yield in Recon-2. 

ATP-coupled reactions are a common, though not the only, source of thermodynamic inconsistencies that can be spotted in Recon-2 (see Supporting File 2 for the complete list of cycles we identified in the Recon-2-derived cell-type specific networks). It is however important to stress that they are spurious and may be identified easily by complementing Recon-2 with a maintenance reaction that mimics the energy expenditure associated with basal processes (similar to those that are present in bacterial metabolic networks), and even cured automatically (or with an automated procedure) by fixing the directionality of active transports directly in the reconstruction (when possible).

\subsection{Correcting infeasible loops in FBA solutions for cell-type specific human metabolic networks}

In this section, we focus on finding and correcting infeasible loops in FBA solutions of the cell-type specific human metabolic networks obtained by Recon-2. We have restricted our attention to 15 networks carrying an objective function, representing respectively cerebral cortex neuronal cell, liver bile duct cell, cervix uterine squamous epithelial cell, kidney tubule cell, gall bladder cell, lung macrophage, small intestine glandular cell, rectum glandular cell, smooth muscle cell, urinary bladder urothelial cell, pre- and post-menopause uterus glandular cell, pancreatic exocrine glandular cell, tonsil germinal cell and squamous epithelial cell. We first computed the FBA solutions for each of the networks via the COBRA Toolbox \cite{cobra2}\footnote{We have computed optimal solutions with respect to the biomass objective function. The choice is mainly motivated by the fact that maximizing the biomass yield represents a network-wide goal with respect to the more specific tasks described by other objective functions included in the reconstructions. We stress however that for our present purposes the obective function merely provides a means of obtaining flux patterns, hence the particular choice we made is immaterial for the problem we consider.}. Subsequently, we identified infeasible loops using the method described in Section \ref{monte} and, finally, corrected thermodynamic inconsistencies using both the local and global strategies described in Sections \ref{correct1} and \ref{correct2} (in the latter case, minimizing the $Q_1$ norm, while fixing sinks, uptakes, and objective function to the values of the FBA solution.) In particular, with the local strategy we eliminate one infeasible loop at a time making sure that no constraint is violated by the corrected solutions, including the value of the objective function. We note however that the local strategy does not return a unique thermodynamically consistent network, since the final flux pattern may depend on the order with which loops are removed. 
We shall see that, quite generically, this strategy produces flux patterns that are more similar to the original (infeasible) solutions than those generated by the global correction strategy.

Results are shown in Table~\ref{TableFBA}, where we list different topological quantities (specifically, the overall number of reactions and metabolites, and the number of reactions carrying a non-zero flux and that of metabolites that are produced and consumed by at least one reaction) for the original (infeasible) FBA solutions and for the corrected flux patterns, both for the local and global strategies, as well as the number of loops in the FBA solutions that need to be corrected by the local strategy. One sees that the local strategy typically needs to resolve several hundreds of inconsistencies in order to obtain viable solutions, and that correction strategies enforce a reduction in the number of active processes, that in certain cases can be rather dramatic. Supporting File 2 lists the cycles we identified and corrected in each of the 15 metabolic networks we have analyzed. To quantify more precisely the similarity between the solutions thus obtained, we have measured the `overlap' parameter defined as follows: given two flux configurations $\mathbf{v}^a=\{v_r^a\}$ and $\mathbf{v}^b=\{v_r^b\}$, we let
\begin{equation}\label{ovle}
q_{ab}=\frac{2}{N}\sum_{r=1}^N\frac{v^a_r v^b_r}{(v_r^a)^2+(v_r^b)^2}~~.
\end{equation}
Clearly, $q_{ab}=1$ if $\mathbf{v}^a=\mathbf{v}^b$, while the more different fluxes are in the two solutions the smaller $q_{ab}$ gets, until $q_{ab}=-1$ if $\mathbf{v}^a=-\mathbf{v}^b$. Larger values of $q_{ab}$ therefore generically point to the fact that the two flux vectors are more similar also in terms of their directions. (Note that in computing (\ref{ovle}) one should account for the fact that a flux that is null in both solutions contributes 1 to the above sum \footnote{In this study, for numerical reasons a flux $v_r$ is considered to be null whenever $v_r<v_0$, where $v_0$ is a (small) threshold. Results have been obtained with $v_0=10^{-6}$ but they are robust to changes in this value.}.) Values of the overlaps between the three solutions we consider (original FBA, FBA corrected by the local strategy, FBA corrected by the global strategy) are also displayed in Table \ref{TableFBA}, clarifying that the local strategy applied to our sample always generates flux patterns that are closer to the original (infeasible) solution than those obtained by the global strategy. Nevertheless, the overlap between the locally- and globally-corrected solutions can also be rather large in some cases, suggesting that a common physical, possibly variational, requirement may underlie, to some extent, the two criteria.

The final column of Table \ref{TableFBA} shows the sign of the Gibbs energy change of ATP hydrolysis that is obtained in the solution corrected by the global strategy. This provides an interesting check of physiologic consistency, as solutions should be compatible with a spontaneous ATP hydrolysis in vivo (i.e. with a negative Gibbs energy difference). We find that only for five models does $Q_1$ minimization provide (thermodynamically feasible) flux configurations carrying a negative Gibbs energy difference for ATP hydrolysis. A possible, simple to obtain improvement of the method we present indeed includes taking into account physiological aspects when correcting a flux configuration. We stress once more, however, that these types of infeasibilities are due to inconsistent constraints or wrong reversibility assignments that prevent the existence of feasible, energetically realistic flux patterns, and can be eliminated already at the stage of network reconstruction. Our main goal here was to show that our method is capable of identifying and correcting loops. By this type of examples we prove that it can furthermore point to possible limitations of the current models.

\section{Discussion}

Accounting for thermodynamic constraints in stoichiometry-based flux models, though potentially highly rewarding (in terms of the possibility to predict metabolite levels, chemical potentials, reaction free energies and reversibility), is a generically hard task. Methods that integrate directly with the constraints defining the space of viable fluxes are often computationally intensive and either presuppose prior biochemical knowledge or lead to a considerable increase in the number of parameters (or both). The technique presented here makes use of stoichiometry alone (hence, it is essentially a topological method) and allows to accomplish two goals: on one hand, counting and listing the infeasible reaction cycles that spur flux configurations derived from thermodynamics-free models; on the other, correcting such infeasibilities in a physically motivated manner. Indeed, we have first analyzed the genome scale metabolic network reconstruction iAf1260 of the bacterium {\it E. coli}. By simply recording the cycles found in randomly generated flux patterns we are able to uncover a much larger set of (much more complex) loops than previously obtained, also involving a much larger overall number of processes, comparing in particular with \cite{wagner} (in this sense outperforming previously employed methods). In passing, we note that our method comes with a certificate of completeness for the set of cycles, which was previously unavailable. Secondly, after showing that cycles plague FBA solutions for the metabolic networks of several different types of human cells (all retrieved from the human Recon-2 reactome), we have applied our loop-removal strategies in order to obtain thermodynamically viable flux patterns that both preserve the basic constraints of FBA as well as the value of the objective function. In doing so, some inconsistencies in the reconstructions have been identified, that can easily be eliminated at the level of network building. Quite importantly in our view, we have also discussed the possibility to employ global variational criteria to generate thermodynamically feasible flux configurations. In particular, generalizing a previous observation, we have proved that flux patterns that minimize the $p$-norms of the fluxes are thermodynamically viable, provided they are allowed by the constraints. Otherwise, this idea can be used (with some care) to remove cycles that do not involve reactions that cannot be inverted or silenced.

The work presented here extends and improves over previous studies, and takes several steps to suggest controlled and motivated methods to deal with thermodynamic inconsistencies in large networks of biochemical reactions. Further improvements along the lines discussed above (requiring e.g. more precise physiological constraints) are clearly possible. Most promisingly, however, we believe that work directed at enhancing the integration of thermodynamic constraints into flux analysis would be extremely important in light of the current efforts aimed at increasing the scope, reach and predictive power of computational models of cellular metabolism. In absence of sufficiently detailed biochemical information about metabolite levels {\it in vivo} or chemical potentials, general stoichiometry-based techniques must be expected to play a key role in this endeavour.


~

{\it Acknowledgments.} This work is supported by the DREAM Seed Project of the Italian Institute of Technology (IIT). The IIT Platform Computation is gratefully acknowledged.


\begin{table*}
\begin{tabular}{  | c  | c  |  c | c | c | c | c | c | c | c | c | c | c | c |}  
\hline
Cell type &  $N$ & $M$ & $N_{FBA}$ &    $M_{FBA}$  &	\# cycles	&   $N_{local}$ & $M_{local}$ &  $N_{global}$ & $M_{global}$  & $q_{FBA,local}$ & $q_{FBA,global}$ &  $q_{local,global}$ & $\Delta$G sign \\
\hline
Bile duct     & 2076 & 1445 & 1009 & 743  & 215 & 516  & 554  & 367 & 476 & 0.706 & 0.559 & 0.781 & + \\
Cer. cortex   & 2169 & 1494 & 1231 & 898  & 358 & 818  & 767  & 257 & 320 & 0.750 & 0.448 & 0.629 & + \\
Cerv. uterine & 1774 & 1171 & 1046 & 780  & 194 & 562  & 620  & 339 & 380 & 0.666 & 0.480 & 0.735 & - \\
Gall bladder  & 3073 & 2159 & 1666 & 1284 & 385 & 1514 & 1227 & 254 & 356 & 0.751 & 0.471 & 0.521 & + \\
Kidney        & 3176 & 2212 & 1695 & 1285 & 414 & 1423 & 1196 & 142 & 449 & 0.759 & 0.469 & 0.551 & + \\
Lung macroph. & 2810 & 1991 & 1313 & 960  & 223 & 817  & 779  & 606 & 587 & 0.765 & 0.681 & 0.849 & - \\
Pancreas      & 2821 & 1951 & 1319 & 948  & 409 & 814  & 797  & 225 & 534 & 0.756 & 0.534 & 0.701 & + \\
Rectum        & 2976 & 2041 & 1328 & 1135 & 406 & 989  & 1017 & 259 & 399 & 0.765 & 0.560 & 0.670 & - \\
Small intest. & 3179 & 2213 & 1385 & 1192 & 405 & 836  & 1023 & 185 & 206 & 0.776 & 0.578 & 0.745 & + \\ 
Smooth muscle & 1806 & 1222 & 1042 & 796  & 184 & 579  & 607  & 314 & 320 & 0.677 & 0.501 & 0.747 & + \\
Tonsil ger.   & 2126 & 1421 & 1178 & 884  & 405 & 881  & 764  & 357 & 412 & 0.667 & 0.503 & 0.644 & - \\
Tonsil sq.    & 2573 & 1718 & 1719 & 1250 & 423 & 1455 & 1188 & 301 & 403 & 0.718 & 0.334 & 0.430 & + \\
Ur. bladder   & 2874 & 1965 & 1597 & 1308 & 219 & 1111 & 1158 & 148 & 686 & 0.760 & 0.450 & 0.613 & + \\
Ut. post-m.   & 2773 & 1973 & 1266 & 1095 & 305 & 736  & 927  & 303 & 389 & 0.763 & 0.578 & 0.757 & + \\
Ut. pre-m.    & 2793 & 1982 & 1376 & 1157 & 208 & 924  & 1022 & 259 & 582 & 0.785 & 0.507 & 0.658 & + \\
\hline
\end{tabular}
\caption{Overview of results obtained for the human tissue specific metabolic networks (with biomass objective function). Columns are as follows. $N$ and $M$: overall number of reactions and metabolites appearing in the network. $N_{FBA}$ and $M_{FBA}$: number of active reactions and produced/consumed metabolites in the FBA solution. \# cycles: number of cycles that the local strategy needs to correct. $N_{local}$ and $M_{local}$: number of active reactions and produced/consumed metabolites in the FBA solution corrected by the local strategy. $N_{global}$ and $M_{global}$: the number of active reactions and produced/consumed metabolites in the FBA solution corrected by the global strategy. $q_{FBA,local}$: overlap between the FBA solution and the solution corrected by the local strategy. $q_{FBA,global}$: overlap between the FBA solution and the solution corrected by the global strategy. $q_{local,global}$: overlap between the FBA solution corrected by the local and global strategies. $\Delta$G sign:  sign of the free energy difference obtained for the ATP hydrolysis in the solution obtained via the global correction strategy.
}
\label{TableFBA}
\end{table*}

 

\bibliography{metabib.bib}

\providecommand{\noopsort}[1]{}\providecommand{\singleletter}[1]{#1}%
\begin{thebibliography}{39}
\providecommand{\url}[1]{\texttt{#1}}
\providecommand{\urlprefix}{ }

\bibitem[Price et~al.(2002)Price, Famili, Beard, and Palsson]{Price:2002mw}
Price, N., I.~Famili, D.~Beard, and B.~Palsson, 2002.
\newblock Extreme Pathways and Kirchhoff's Second Law.
\newblock \emph{Biophys. J.} 83:2879.

\bibitem[Soh and Hatzimanikatis(2010)]{nethatzi}
Soh, K., and V.~Hatzimanikatis, 2010.
\newblock Network thermodynamics in the post-genomic era.
\newblock \emph{Curr. Opin. Microbiol.} 13:350.

\bibitem[Beard et~al.(2004)Beard, Babson, Curtis, and Qian]{babson}
Beard, D., E.~Babson, E.~Curtis, and H.~Qian, 2004.
\newblock Thermodynamic constraints for biochemical networks.
\newblock \emph{J. Theor. Biology} 228:327.

\bibitem[Hoppe et~al.(2007)Hoppe, Hoffmann, and Holzhutter]{Hoppe:2007cr}
Hoppe, A., S.~Hoffmann, and H.~Holzhutter, 2007.
\newblock Including metabolite concentrations into flux balance analysis:
  thermodynamic realizability as a constraint on flux distributions in
  metabolic networks.
\newblock \emph{BMC Systems Biology} 1:23.

\bibitem[Qian and Beard(2005)]{Qian:2005ve}
Qian, H., and D.~Beard, 2005.
\newblock Thermodynamics of stoichiometric biochemical networks in living
  systems far from equilibrium.
\newblock \emph{Biophys. Chem.} 114:213.

\bibitem[Beard et~al.(2002)Beard, Liang, and Qian]{Beard:2002vn}
Beard, D., S.~Liang, and H.~Qian, 2002.
\newblock Energy Balance for Analysis of Complex Metabolic Networks.
\newblock \emph{Biophys. J.} 83:79.

\bibitem[Palsson(2006)]{palssonbook}
Palsson, B.~O., 2006.
\newblock Systems Biology: Properties of Reconstructed Networks.
\newblock Cambridge University Press.

\bibitem[Bowden(2013)]{AthelCBbook}
Bowden, A.~C., 2013.
\newblock Fundamentals of enzyme kinetics.
\newblock Wiley-Blackwell.

\bibitem[Ge et~al.(2012)Ge, Qian, and Qian]{stocqian}
Ge, H., M.~Qian, and H.~Qian, 2012.
\newblock Stochastic theory of nonequilibrium steady states. Part II:
  Applications in chemical biophysics.
\newblock \emph{Phys. Rep.} 510:87.

\bibitem[Frey and Kroy(2005)]{brownianbio}
Frey, E., and K.~Kroy, 2005.
\newblock Brownian motion: a paradigm of soft matter and biological physics.
\newblock \emph{Annalen der Physik} 14:20.

\bibitem[Beg et~al.(2007)Beg, Vazquez, Ernst, de~Menezes, Bar-Joseph,
  Barab{\'a}si, and Oltvai]{Beg:2007mq}
Beg, Q., A.~Vazquez, J.~Ernst, M.~de~Menezes, Z.~Bar-Joseph, A.-L.
  Barab{\'a}si, and Z.-N. Oltvai, 2007.
\newblock Intracellular crowding defines the mode and sequence of substrate
  uptake by Escherichia coli and constrains its metabolic activity.
\newblock \emph{Proc. Nat. Acad. Sci. USA} 104:12663.

\bibitem[{De Martino} and Marinari(2010)]{Kyoto10}
{De Martino}, A., and E.~Marinari, 2010.
\newblock The solution space of metabolic networks: producibility, robustness
  and fluctuations.
\newblock \emph{J. Phys. Conf. Ser.} 233:012019.

\bibitem[Schrijver(1986)]{linearpro}
Schrijver, A., 1986.
\newblock Theory of linear and integer programming.
\newblock Wiley.

\bibitem[Orth et~al.(2010)Orth, Thiele, and Palsson]{Orth:2010if}
Orth, J., I.~Thiele, and B.-O. Palsson, 2010.
\newblock What is flux balance analysis?
\newblock \emph{Nature Biotechnol.} 28:245.

\bibitem[Segr{\`e} et~al.(2002)Segr{\`e}, Vitkup, and Church]{Segre:2002oq}
Segr{\`e}, D., D.~Vitkup, and G.~Church, 2002.
\newblock Analysis of optimality in natural and perturbed metabolic networks.
\newblock \emph{Proc. Nat. Acad. Sci. USA} 99:15112.

\bibitem[Jankowski et~al.(2008)Jankowski, Henry, Broadbelt, and
  Hatzimanikatis]{Jankowski:2008bh}
Jankowski, M., C.~Henry, L.~Broadbelt, and V.~Hatzimanikatis, 2008.
\newblock Group contribution method for thermodynamic analysis of complex
  metabolic networks.
\newblock \emph{Biophys. J.} 95:1487.

\bibitem[Fleming et~al.(2009)Fleming, Thiele, and Nasheuer]{Fleming:2009qa}
Fleming, R., I.~Thiele, and H.~Nasheuer, 2009.
\newblock Quantitative assignment of reaction directionality in
  constraint-based models of metabolism: Application to Escherichia coli.
\newblock \emph{Biophys. Chem.} 145:47.

\bibitem[Kummel et~al.(2006)Kummel, Panke, and Heinemann]{Kummel:2006bh}
Kummel, A., S.~Panke, and M.~Heinemann, 2006.
\newblock Systematic assignment of thermodynamic constraints in metabolic
  network models.
\newblock \emph{BMC Bioinformatics} 7:512.

\bibitem[Alberty(2003)]{Albertybook}
Alberty, R.~A., 2003.
\newblock Thermodynamics of Biochemical Reactions.
\newblock Wiley.

\bibitem[Schellenberger et~al.(2011{\natexlab{a}})Schellenberger, Lewis, and
  Palsson]{Schellenberger:2011uq}
Schellenberger, J., N.~Lewis, and B.-O. Palsson, 2011.
\newblock Elimination of thermodynamically infeasible loops in steady-state
  metabolic models.
\newblock \emph{Biophys. J.} 100:544.

\bibitem[Henry et~al.(2007)Henry, Broadbelt, and Hatzimanikatis]{Henry:2007xz}
Henry, C., L.~Broadbelt, and V.~Hatzimanikatis, 2007.
\newblock Thermodynamics-Based Metabolic Flux Analysis.
\newblock \emph{Biophys. J.} 92:1792.

\bibitem[M\"uller and Brockmayr(2013)]{fastthermo}
M\"uller, A., and A.~Brockmayr, 2013.
\newblock Fast thermodynamically constrained flux variability analysis.
\newblock \emph{Bioinformatics} 29:903.

\bibitem[Beard and Qian(2008)]{chembiop}
Beard, D.-A., and H.~Qian, 2008.
\newblock Chemical biophysics.
\newblock Cambridge University Press.

\bibitem[{De Martino} et~al.(2012{\natexlab{a}}){De Martino}, Figliuzzi, {De
  Martino}, and Marinari]{noiplos}
{De Martino}, D., M.~Figliuzzi, A.~{De Martino}, and E.~Marinari, 2012.
\newblock A Scalable Algorithm to Explore the Gibbs energy Landscape of
  Genome-scale Metabolic Networks.
\newblock \emph{PLoS Comp. Biol.} 8:e1002562.

\bibitem[{De Martino}(2013)]{deMartino12}
{De Martino}, D., 2013.
\newblock Thermodynamics of biochemical networks and duality theorems.
\newblock \emph{Phys. Rev. E} 87:053108.

\bibitem[Johnson(1975)]{jnsn}
Johnson, D.-B., 1975.
\newblock Finding all the elemtary circuits of a directed graph.
\newblock \emph{SIAM J. on Computing} 4:77.

\bibitem[Wright and Wagner(2008)]{wagner}
Wright, J., and A.~Wagner, 2008.
\newblock Exhaustive identification of steady state cycles in large
  stoichiometric networks.
\newblock \emph{BMC Systems Biology} 2:61.

\bibitem[Wiback et~al.(2004)Wiback, Famili, Greenberg, and
  Palsson]{Wiback:2004kc}
Wiback, S., I.~Famili, H.~Greenberg, and B.-O. Palsson, 2004.
\newblock Monte Carlo sampling can be used to determine the size and shape of
  the steady-state flux space.
\newblock \emph{J. Theor. Biol.} 228:437.

\bibitem[Price et~al.(2004)Price, Schellenberger, and Palsson]{Price:2004lp}
Price, N., J.~Schellenberger, and B.-O. Palsson, 2004.
\newblock Uniform Sampling of Steady-State Flux Spaces: Means to Design
  Experiments and to Interpret Enzymopathies.
\newblock \emph{Biophys. J.} 87:2172.

\bibitem[Mezard and Montanari(2009)]{mezmon}
Mezard, M., and A.~Montanari, 2009.
\newblock Information, Physics, and Computation.
\newblock Oxford University Press.

\bibitem[Feist et~al.(2007)Feist, Henry, Reed, Krummenacker, Joyce, Karp,
  Broadbelt, Hatzimanikatis, and Palsson]{Feist:2007zk}
Feist, A., C.~Henry, J.~Reed, M.~Krummenacker, A.~Joyce, P.~Karp, L.~Broadbelt,
  V.~Hatzimanikatis, and B.-O. Palsson, 2007.
\newblock A genome-scale metabolic reconstruction for Escherichia coli K-12
  MG1655 that accounts for 1260 ORFs and thermodynamic information.
\newblock \emph{Mol. Sys. Biol.} 3:121.

\bibitem[Thiele et~al.(2013)Thiele, Swainston, Fleming, Hoppe, Sahoo, Aurich,
  Haraldsdottir, Mo, Rolfsson, Stobbe, Thorleifsson, Agren, B{\"o}lling,
  Bordel, Chavali, Dobson, Dunn, Endler, Hala, Hucka, Hull, Jameson, Jamshidi,
  Jonsson, Juty, Keating, Nookaew, Nov{\`e}re, Malys, Mazein, Papin, Price,
  Selkov, Sigurdsson, Simeonidis, Sonnenschein, Smallbone, Sorokin, van Beek,
  Weichart, Goryanin, Nielsen, Westerhoff, Kell, Mendes, and Palsson]{Recon2}
Thiele, I., N.~Swainston, R.~M.~T. Fleming, A.~Hoppe, S.~Sahoo, M.~K. Aurich,
  H.~Haraldsdottir, M.~L. Mo, O.~Rolfsson, M.~D. Stobbe, S.~G. Thorleifsson,
  R.~Agren, C.~B{\"o}lling, S.~Bordel, A.~K. Chavali, P.~Dobson, W.~B. Dunn,
  L.~Endler, D.~Hala, M.~Hucka, D.~Hull, D.~Jameson, N.~Jamshidi, J.~J.
  Jonsson, N.~Juty, S.~Keating, I.~Nookaew, N.~L. Nov{\`e}re, N.~Malys,
  A.~Mazein, J.~A. Papin, N.~D. Price, E.~Selkov, M.~I. Sigurdsson,
  E.~Simeonidis, N.~Sonnenschein, K.~Smallbone, A.~Sorokin, J.~H. G.~M. van
  Beek, D.~Weichart, I.~Goryanin, J.~Nielsen, H.~V. Westerhoff, D.~B. Kell,
  P.~Mendes, and B.~{\O}. Palsson, 2013.
\newblock A community-driven global reconstruction of human metabolism.
\newblock \emph{Nature Biotechnol.} 31:419.

\bibitem[Schellenberger et~al.(2011{\natexlab{b}})Schellenberger, Que, Fleming,
  Thiele, Orth, Feist, Zielinski, Bordbar, Lewis, Rahmanian, Kang, Hyduke, and
  Palsson]{cobra2}
Schellenberger, J., R.~Que, R.~M.~T. Fleming, I.~Thiele, J.~D. Orth, A.~M.
  Feist, D.~C. Zielinski, A.~Bordbar, N.~E. Lewis, S.~Rahmanian, J.~Kang, D.~R.
  Hyduke, and B.-O. Palsson, 2011.
\newblock Quantitative prediction of cellular metabolism with constraint-based
  models: the COBRA Toolbox v2.0.
\newblock \emph{Nature Protocols} 6:1290.

\bibitem[Shlomi et~al.(2008)Shlomi, Cabili, Herrg{\aa}rd, Palsson, and
  Ruppin]{shlomi2008}
Shlomi, T., M.~N. Cabili, M.~J. Herrg{\aa}rd, B.-O. Palsson, and E.~Ruppin,
  2008.
\newblock Network-based prediction of human tissue-specific metabolism.
\newblock \emph{Nature Biotechnol.} 26:1003.

\bibitem[Krauth and Mezard(1987)]{minover0}
Krauth, W., and M.~Mezard, 1987.
\newblock Learning algorithms with optimal stability in neural networks.
\newblock \emph{J. Phys. A: Math. Gen.} 20:L745.

\bibitem[Binder and Heermann(2002)]{montec}
Binder, K., and D.-W. Heermann, 2002.
\newblock Monte Carlo Simulation in Statistical Physics.
\newblock Springers.

\bibitem[{De Martino} et~al.(2012{\natexlab{b}}){De Martino}, {De Martino},
  Mulet, and Uguzzoni]{plos1}
{De Martino}, A., D.~{De Martino}, R.~Mulet, and G.~Uguzzoni, 2012.
\newblock Reaction Networks as Systems for Resource Allocation: A Variational
  Principle for Their Non-Equilibrium Steady States.
\newblock \emph{PLoS ONE} 7:e39849.

\bibitem[Schilling et~al.(2000)Schilling, Letscher, and
  Palsson]{Schilling:2000p4106}
Schilling, C.-H., D.~Letscher, and B.-O. Palsson, 2000.
\newblock Theory for the systemic definition of metabolic pathways and their
  use in interpreting metabolic function from a pathway-oriented perspective.
\newblock \emph{J. Theor. Biol.} 203:229.

\bibitem[Mahadevan and Schilling(2003)]{Mahadevan:2003pi}
Mahadevan, R., and C.~Schilling, 2003.
\newblock The effects of alternate optimal solutions in constraint-based
  genome-scale metabolic models.
\newblock \emph{Metab. Eng.} 5:264.

\end{thebibliography}
\bibliographystyle{biophysj}

\end{document}